\def\rfr#1{eq. (\ref{#1})}

\def\cf#1#2{\dot\Omega^{\rm #2}_{.#1}}

\def\dert#1#2{\frac{{{d}}{#1}}{{{d}}{#2}}}              

\def\bar{\begin{eqnarray}}
\def\ear{\end{eqnarray}}
\def\bb{\bibitem}
\def\eqi{\begin{equation}}
\def\eqf{\end{equation}}
\def\eqia{\begin{eqnarray}}
\def\eqfa{\end{eqnarray}}
\def\rp#1#2{{#1\over#2}}

\def\lb#1{\label{#1}}

\def\oc2{$\mathcal{O}(c^{-2})$}

\def\bds#1{\vec{\it{#1}}}

\documentclass[natbib]{svjour3}                     
\smartqed  
\usepackage{graphicx}
 \usepackage{aps-bibstyle}  
\begin{document}

\title{An Assessment of the Systematic Uncertainty in Present and Future Tests of the Lense-Thirring Effect with Satellite Laser Ranging
}

\titlerunning{On the Systematic Uncertainty in Present and Future Tests of the Lense-Thirring Effect with SLR}        

\author{Lorenzo Iorio        }

\authorrunning{Lorenzo Iorio} 

\institute{L. Iorio \at
              INFN-Sezione di Pisa \\
              Tel.: +39-328-6128815\\
              \email{lorenzo.iorio@libero.it}            \\
             \emph{Present address:}  Viale Unit\`{a} di Italia 68, 70125, Bari (BA), Italy 
           }

\date{Received: date / Accepted: date}

\maketitle

\begin{abstract}
We deal with the attempts to measure  the Lense-Thirring effect with the Satellite Laser Ranging (SLR) technique applied to the existing LAGEOS and LAGEOS II terrestrial satellites and to the recently approved LARES spacecraft. According to general relativity, a central spinning body of mass $M$ and angular momentum $\bds S$ like the Earth generates a gravitomagnetic field which induces small secular precessions of the orbit of a test particle geodesically moving around it.
Extracting this signature from the data is a demanding task because of many classical orbital perturbations having the same pattern as the gravitomagnetic one, like those due to the centrifugal oblateness of the Earth which represents a major source of systematic bias.
The first issue addressed here is: are the so far published evaluations of the systematic uncertainty induced by  the bad knowledge of the even zonal harmonic coefficients $J_{\ell}$ of the multipolar expansion of the Earth's geopotential  reliable and realistic?
 Our answer is negative. Indeed, if the differences $\Delta J_{\ell}$ among the even zonals estimated in different Earth's gravity field global solutions from the dedicated GRACE mission are assumed for the uncertainties $\delta J_{\ell}$  instead of using their covariance sigmas $\sigma_{J_\ell}$, it turns out that the systematic uncertainty $\delta\mu$ in the Lense-Thirring test with the nodes $\Omega$ of LAGEOS and LAGEOS II may be up to 3 to 4 times larger than in the evaluations so far published ($5-10\%$) based on the use of the sigmas of one model at a time separately.  The second issue consists of the possibility of using a different approach in extracting the relativistic signature of interest from the LAGEOS-type data. The third issue is the possibility of reaching a realistic total accuracy of $1\%$ with LAGEOS, LAGEOS II and LARES, which should be launched in November 2009 with a VEGA rocket.   While LAGEOS and LAGEOS II fly at altitudes of about 6000 km, LARES will be likely placed at an altitude of 1450 km. Thus, it will be sensitive to much more even zonals than LAGEOS and LAGEOS II. Their corrupting impact has been evaluated with the standard Kaula's approach up to degree $\ell=60$ by using $\Delta J_{\ell}$ and $\sigma_{J_{\ell}}$; it turns out that it may be as large as some tens percent. The different orbit of LARES may also have some consequences on the non-gravitational orbital perturbations affecting it which might further degrade the obtainable accuracy.
\end{abstract}

 \section{Introduction}
In the weak-field and slow motion approximation,  the Einstein field equations of general relativity get linearized resembling to the Maxwellian equations of electromagntism. As a consequence, a gravitomagnetic field $\bds B_{\rm g}$, induced by the off-diagonal components $g_{0i}, i=1,2,3$ of the space-time metric tensor related to the mass-energy currents of the source of the gravitational field, arises \citep{MashNOVA}. The gravitomagnetic field affects orbiting test particles, precessing gyroscopes, moving clocks and atoms and propagating electromagnetic waves \citep{Rug,Scia04}. Perhaps, the most famous gravitomagnetic effects are the precession  of the axis of a gyroscope \citep{Pugh,Schi} and the Lense-Thirring\footnote{According to an interesting historical analysis recently performed by \citet{Pfi07}, it would be more correct to speak about an Einstein-Thirring-Lense effect.} precessions \citep{LT} of the orbit of a test particle, both occurring in the field of a central slowly rotating mass like, e.g., our planet.   Direct, undisputable measurements of such  fundamental predictions of general relativity are not yet available.

The measurement of the gyroscope precession in the Earth's gravitational field has been the goal of the dedicated space-based\footnote{See on the WEB http://einstein.stanford.edu/} GP-B mission \citep{Eve, GPB} launched in 2004 and carrying onboard four superconducting gyroscopes; its data analysis is still ongoing. The target accuracy was originally $1\%$, but it is still unclear if the GP-B team will succeed in reaching such a goal because of some unmodelled effects affecting the gyroscopes: 1) a time variation in the polhode motion of the gyroscopes and 2) very large classical misalignment torques on the gyroscopes.

\begin{figure}
   \caption{Keplerian orbit. The longitude of the ascending node $\Omega$ is counted from a reference X direction  in the equator of the central body, assumed as reference plane $\{{\rm X,Y}\}$,  to the line of the nodes which is the intersection of the orbital plane with the equatorial plane of the central body. It has mass $M$ and proper angular momentum $\bds S$. The argument of pericentre $\omega$ is an angle in the orbital plane counted from the line of the nodes to the location of the pericentre, here marked with $\Pi$. The time-dependent position of the moving test particle of mass $m$ is given by the true anomaly $f$, counted anticlockwise from the pericentre's position. The inclination between the orbital and the equatorial planes is $i$. Thus, $\Omega,\omega, i$ can be viewed as the three (constant) Euler angles fixing the configuration of a rigid body, i.e. the orbit which in the unperturbed Keplerian case does  change neither its shape nor its size, in the inertial $\{{\rm X,Y,Z}\}$ space.
   Courtesy by H.I.M. Lichtenegger, IWF, Graz.}
   \label{plo}
   \includegraphics{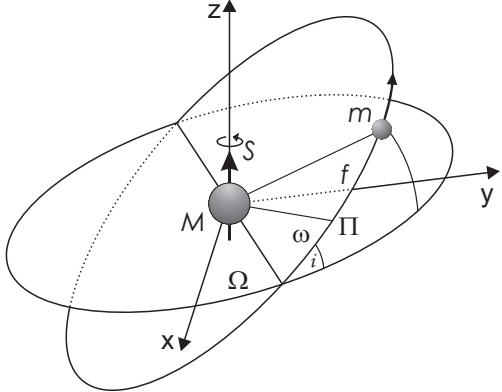}
   \end{figure}

In this Chapter we  will focus on the attempts to measure of the Lense-Thirring effect in the gravitational field of the Earth; for Mars and the Sun see \citep{LTmars,LTkrogh,LTreplytokrogh} and \citep{LTsun}, respectively.
Far from a localized rotating body  with angular momentum ${\bds S}$  the gravitomagnetic field can be written as
\eqi \bds B_{\rm g} = -\rp{G}{c r^3}\left[\bds S -3\left(\bds S\bds\cdot{\hat{r}}\right){\hat{r}}\right],\eqf   where $G$ is the Newtonian gravitational constant and $c$ is the speed of light in vacuum. It acts on a test particle orbiting with a velocity $\bds v$ with the non-central acceleration \citep{Sof}
\eqi \bds A_{\rm LT} = -\rp{2}{c}{\bds v}\bds\times\bds B_{\rm g}\eqf
which induces secular precessions of the longitude of the ascending node $\Omega$ \begin{equation}\dot\Omega_{\rm LT} = \rp{2 G S}{c^2 a^3 (1-e^2)^{3/2}},\lb{let}\end{equation}
and the argument of pericentre $\omega$
\begin{equation}\dot\omega_{\rm LT} = -\rp{6 G S\cos i}{c^2 a^3 (1-e^2)^{3/2}},\lb{LT_o}\end{equation}
of the orbit of  a test particle. In \rfr{let} and \rfr{LT_o}
 $a$ and $e$ are the semimajor axis and the eccentricity, respectively, of the test particle's orbit and $i$ is its inclination to the central body's equator.
The semimajor axis $a$ fixes the size of the ellipse, while its shape is determined by the eccentricity $0\leq e<1$; an orbit with $e=0$ is a circle. The angles $\Omega$ and $\omega$ establish the orientation of the orbit in the inertial space and in the orbital plane, respectively. $\Omega$, $\omega$ and $i$ can be viewed as the three Euler angles which determine the orientation of a rigid body with respect to an inertial frame. In Figure \ref{plo} we illustrate the geometry of a Keplerian orbit.

In this Chapter we will critically discuss the following  topics
\begin{itemize}
  \item Section \ref{grav}. The realistic evaluation  of the total accuracy in the test performed in recent years with the existing Earth's artificial satellites
  LAGEOS and LAGEOS II \citep{Ciu04,Ciu06,Ries08}. LAGEOS was put into orbit in 1976, followed by its twin LAGEOS II in 1992; they are passive, spherical spacecraft entirely covered by retroreflectors which allow for their accurate tracking through laser pulses sent from Earth-based ground stations according to the Satellite Laser Ranging (SLR) technique \citep{slr}. They orbit at altitudes of about 6000 km ($a_{\rm LAGEOS} =12270$ km, $a_{\rm LAGEOS\ II}=12163$ km) in nearly circular paths ($e_{\rm LAGEOS}=0.0045$, $e_{\rm LAGEOS\ II}=0.014$) inclined by 110 deg and 52.65 deg, respectively, to the Earth's equator. The Lense-Thirring effect for their nodes amounts to about 30 milliarcseconds per year (mas yr$^{-1}$) which correspond to about 1.7 m yr$^{-1}$ in the cross-track direction\footnote{A perturbing acceleration like $\bds A_{\rm LT}$ is customarily projected onto the radial $\bds{\hat r}$, transverse $\bds{\hat \tau}$ and cross-track $\bds{\hat \nu}$ directions of an orthogonal  frame comoving with the satellite \citep{Sof}; it turns out that the Lense-Thirring node precession affects the cross-track component of the orbit according to $\Delta\nu_{\rm LT} \approx a\sin i\Delta\Omega_{\rm LT}$ (eq. (A65), p. 6233 in \citep{Cri}).} at the LAGEOS altitudes.

  The idea of measuring the Lense-Thirring node rate with the just launched LAGEOS satellite, along with the other SLR targets orbiting at that time, was put forth by \citet{Cug78}. Tests have started to be effectively performed later by using the LAGEOS and LAGEOS II satellites \citep{tanti}, according to a strategy  by \citet{Ciu96} involving the use of a suitable linear combination of the nodes $\Omega$ of both satellites and the perigee $\omega$ of LAGEOS II. This was done to reduce the impact of the most relevant source of systematic bias, i.e. the mismodelling in the even ($\ell=2,4,6\ldots$) zonal ($m=0$) harmonic coefficients $J_{\ell}$ of the multipolar expansion of the Newtonian part of the terrestrial gravitational potential due to the diurnal rotation (they induce secular precessions on the node and perigee of a terrestrial satellite much larger than the gravitomagnetic ones. The $J_{\ell}$ coefficients cannot be theoretically computed but must be estimated by suitably processing long data sets from the dedicated satellites like CHAMP and GRACE; see Section \ref{grav}): the three-elements combination used allowed for removing the uncertainties in $J_2$ and $J_4$. In \citep{Ciu98} a $\approx 20\%$ test was reported by using the\footnote{Contrary to the subsequent models based on the dedicated satellites CHAMP (http://www-app2.gfz-potsdam.de/pb1/op/champ/index$\_$CHAMP.html) and GRACE (http://www-app2.gfz-potsdam.de/pb1/op/grace/index$\_$GRACE.html), EGM96 relies upon multidecadal tracking of SLR data of a constellation of geodetic satellites including LAGEOS and LAGEOS II as well; thus the possibility of a sort of $a-priori$ `imprinting' of the Lense-Thirring effect itself, not solved-for in EGM96, cannot be neglected.} EGM96 Earth gravity model \citep{Lem98}; subsequent detailed analyses showed that such an evaluation of the total error budget was overly optimistic in view of the likely unreliable computation of the total bias due to the even zonals \citep{Ior03,Ries03a,Ries03b}.  An analogous, huge underestimation turned out to hold also for the effect of the non-gravitational perturbations \citep{Mil87} like the direct solar radiation pressure, the Earth's albedo, various subtle thermal effects depending on  the physical properties of the satellites' surfaces and their rotational state \citep{Inv94,Ves99,Luc01,Luc02,Luc03,Luc04,Lucetal04,Ries03a}, which the perigees  of LAGEOS-like satellites are particularly sensitive to. As a result, the realistic total error budget in the test reported in \citep{Ciu98} might be as large as $60-90\%$ or (by considering EGM96 only) even more.

The observable used by \citet{Ciu04} with the GRACE-only EIGEN-GRACE02S model \citep{eigengrace02s} and by \citet{Ries08} with other more recent Earth gravity models was the following linear combination\footnote{See also \citep{Pav02,Ries03a,Ries03b}.} of the nodes of LAGEOS and LAGEOS II, explicitly computed by \citet{IorMor} following the approach put forth by \citet{Ciu96}
\begin{equation} f=\dot\Omega^{\rm LAGEOS}+
c_1\dot\Omega^{\rm LAGEOS\ II }, \lb{combi}\end{equation}
where \begin{equation} c_1\equiv-\rp{\dot\Omega^{\rm LAGEOS}_{.2}}{\dot\Omega^{\rm
LAGEOS\ II }_{.2}}=-\rp{\cos i_{\rm LAGEOS}}{\cos i_{\rm LAGEOS\
II}}\left(\rp{1-e^2_{\rm LAGEOS\ II}}{1-e^2_{\rm
LAGEOS}}\right)^2\left(\rp{a_{\rm LAGEOS\ II}}{a_{\rm LAGEOS}}\right)^{7/2}.\lb{coff}\end{equation} The coefficients $\dot\Omega_{.\ell}$ of the aliasing classical node precessions \citep{Kau} $\dot\Omega_{\rm class}=\sum_{\ell}\dot\Omega_{.\ell}J_{\ell}$ induced by the even zonals  have been analytically worked out up to $\ell=20$ in, e.g. \citep{Ior03}; they yield $c_1=0.544$.
 The Lense-Thirring signature of \rfr{combi} amounts to 47.8 mas yr$^{-1}$. The combination  \rfr{combi} allows, by construction, to remove the aliasing effects due to the static and time-varying parts of the first even zonal $J_2$. The nominal (i.e. computed with the estimated values of $J_{\ell}$, $\ell=4,6...$) bias due to the remaining higher degree even zonals would amount to  about $10^5$ mas yr$^{-1}$; the need of a careful and reliable modeling of such an important source of systematic bias is, thus, quite apparent. Conversely, the nodes of the LAGEOS-type spacecraft are directly affected by the non-gravitational accelerations at a $\approx 1\%$ level of the Lense-Thirring effect \citep{Luc01,Luc02,Luc03,Luc04,Lucetal04}. For a comprehensive, up-to-date overview of the numerous and subtle issues concerning the measurement of the Lense-Thirring effect see, e.g., \citep{IorNOVA}.
\item Section \ref{approach}. Another approach which could be followed in extracting the Lense-Thirring effect from the data of the LAGEOS-type satellites.
  \item Section \ref{larez}. The possibility that the LARES mission, recently approved by the Italian Space Agency (ASI), will be able to measure the Lense-Thirring node precession with an accuracy of the order of $1\%$.

      In \citep{vpe76a,vpe76b}  it was proposed to measure the Lense-Thirring precession of the nodes $\Omega$ of a pair of counter-orbiting spacecraft to be launched in terrestrial polar orbits and endowed with drag-free apparatus. A somewhat equivalent, cheaper version of such an idea was put forth in 1986 by \citet{Ciu86} who proposed to launch a passive, geodetic satellite in an orbit identical to that of LAGEOS  apart from the orbital planes which should have been displaced by 180 deg apart. The measurable quantity was, in the case of the proposal by \citet{Ciu86}, the sum of the nodes of LAGEOS and of the new spacecraft, later named LAGEOS III, LARES, WEBER-SAT, in order to cancel to a high level of accuracy the corrupting effect of the multipoles of the Newtonian part of the terrestrial gravitational potential which represent the major source of systematic error (see Section \ref{grav}). Although extensively studied by various groups \citep{CSR,LARES}, such an idea was not implemented for many years.
In \citep{Ioretal02} it was proposed to include also the data from LAGEOS II by using a different observable.  Such an approach was proven in \citep{IorNA} to be potentially useful in making the constraints on the orbital configuration of the new SLR satellite less stringent than it was originally required in view of the recent improvements in our knowledge of the classical part of the terrestrial gravitational potential due to the dedicated CHAMP and, especially, GRACE  missions.

Since reaching high altitudes and minimizing the unavoidable orbital injection errors is expensive, it was explored the possibility of discarding LAGEOS and LAGEOS II using a low-altitude, nearly polar orbit for LARES \citep{LucPao01,Ciu06b}, but in \citep{Ior02,Ior07c} it was proven that such alternative approaches are not feasible. It was also suggested that LARES would be able to probe alternative theories of gravity \citep{Ciu04b}, but also in this case it turned out to be impossible \citep{IorJCAP,Ior07d}.

The stalemate came to an end when ASI recently made the following official announcement  (http://www.asi.it/SiteEN/MotorSearchFullText.aspx?keyw=LARES): ``On February 8, the ASI board approved funding for the LARES mission, that will be launched with VEGA's maiden flight before the end of 2008. LARES is a passive satellite with laser mirrors, and will be used to measure the Lense-Thirring effect.''  The italian version of the announcement yields some more information specifying that LARES, designed in collaboration with National Institute of Nuclear Physics (INFN), is currently under construction by Carlo Gavazzi Space SpA; its Principal Investigator (PI) is I. Ciufolini and its scientific goal is to measure at a $1\%$ level the  Lense-Thirring effect in the gravitational field of the Earth.
Concerning the orbital configuration of LARES,
In one of the latest communication to INFN, Rome, 30 January 2008, \citep{INFN} writes that LARES will be launched with a semimajor axis of approximately 7600 km and an inclination between 60 and 80 deg.
More precise information can be retrieved in Section 5.1, pag 9 of the document Educational Payload on the Vega Maiden Flight
Call For CubeSat Proposals, European Space Agency,
Issue 1
11 February 2008, downloadable at
http://esamultimedia.esa.int/docs/LEX-EC/CubeSat$\%$20CFP$\%$20issue$\%$201.pdf.
    It is  written there that LARES will be launched into a circular orbit with altitude $h=1200$ km, corresponding to a semimajor axis $a_{\rm LARES}=7578$ km, and inclination $i=71$ deg to the Earth's equator. Latest information\footnote{See on the WEB
http://www.esa.int/esapub/bulletin/bulletin135/bul135f$\_$bianchi.pdf.} point towards a launch at the end of 2009 with a VEGA rocket in a circular orbit inclined by 71 deg to the Earth's equator at an altitude of\footnote{I thank Dr. D. Barbagallo (ESRIN) for having kindly provided me with the latest details of the orbital configuration of LARES.} 1450 km corresponding to a semimajor axis of $a_{\rm LR}=7828$ km. More or less the same has been reported by \citet{INFN2} to the INFN in Villa Mondragone, 3 October 2008.
\end{itemize}

\section{The systematic error of gravitational origin in the LAGEOS-LAGEOS II test}\lb{grav}
The realistic evaluation of the total error budget of the LAGEOS-LAGEOS II node test \citep{Ciu04} raised a lively debate
\citep{Ciu05,Ciu06,IorNA,IorJoG,IorGRG,Ior07,Luc05}, mainly focussed on the impact of the static and time-varying parts of the Newtonian component of the Earth's gravitational potential through the secular precessions induced on a satellite's node.

In the real world the path of a probe is not only affected by the relativistic gravitomagentic field but also by a huge number of other competing classical orbital perturbations of gravitational and non-gravitational origin.
The most insidious disturbances are those induced by the static part of the Newtonian component of the multipolar expansion in spherical harmonics\footnote{The relation among the  even zonals $J_{\ell}$ and the  normalized gravity coefficients $\overline{C}_{\ell 0}$ which are customarily determined in the Earth's gravity models, is $J_{\ell}=-\sqrt{2\ell + 1}\ \overline{C}_{\ell 0}$.} $J_{\ell}, \ell = 2,4,6,...$ of the gravitational potential of the central rotating mass \citep{Kau}: they affect the node with effects having the same signature of the relativistic signal of interest, i.e. linear trends which are orders of magnitude larger and cannot be removed from the time series of data without affecting the Lense-Thirring pattern itself as well. The only thing that can be done is to model such a corrupting effect as most accurately as possible and assessing the impact of the residual mismodelling on the measurement of the frame-dragging effect.
The secular precessions induced by the even zonals of the geopotential can be written as
\begin{equation}\dot\Omega^{\rm geopot}=\sum_{\ell  =2}\dot\Omega_{.\ell}J_{\ell},\end{equation}
where the coefficients $\dot\Omega_{.\ell}, \ell=2,4,6,...$ depend on the parameters of the Earth ($GM$ and the equatorial radius $R$) and on the semimajor axis $a$, the inclination $i$ and the eccentricity $e$ of the satellite. For example, for $\ell=2$
we have
\begin{equation}\dot\Omega_{.2}=-\rp{3}{2}n\left(\rp{R}{a}\right)^2\rp{\cos i}{(1-e^2)^2};\end{equation}
 $n=\sqrt{GM/a^3}$ is the Keplerian mean motion.
They have been analytically computed up to $\ell=20$  in, e.g., \citep{Ior03}.
Their mismodelling can be written as
\begin{equation}\delta\dot\Omega^{\rm geopot}\leq \sum_{\ell  =2}\left|\dot\Omega_{.\ell}\right|\delta J_{\ell},\lb{mimo}\end{equation}
where $\delta J_{\ell}$ represents our uncertainty in the knowledge of the even zonals $J_{\ell}$

A common feature of all the competing evaluations  so far published is that the systematic bias due to the static component of  the
  geopotential was always calculated  by using the released (more or less accurately calibrated) sigmas $\sigma_{J_{\ell}}$  of one Earth gravity model solution at a time for the uncertainties $\delta J_{\ell}$. Thus, it was said that the model X yields a $x\%$ error, the model Y   yields a $y\%$ error, and so on.

Since a trustable calibration of the formal, statistical uncertainties in the estimated zonals of the covariance matrix of a global solution is always a difficult task to be implemented in a reliable way, a much more realistic and conservative approach consists, instead, of taking the difference\footnote{See Fig.5 of \citep{Luc07} for a comparison of the estimated $\overline{C}_{40}$ in different models.} \eqi\Delta J_{\ell}=\left|J_{\ell}(\rm X) - J_{\ell}(\rm Y)\right|,\ \ell=2,4,6,...\eqf of the estimated even zonals for different pairs of Earth gravity field solutions as representative of the real uncertainty $\delta J_{\ell}$ in the zonals \citep{Lerch}. In Table \ref{tavola1}--Table \ref{tavolaAIUB2}  we present our results for the most recent GRACE-based models released so far by different institutions and retrievable on the Internet at\footnote{I thank M Watkins (JPL) for having provided me with the even zonals and their sigmas of the  JEM01-RL03B model.}
http://icgem.gfz-potsdam.de/ICGEM/ICGEM.html.   The models used are EIGEN-GRACE02S \citep{eigengrace02s}  from GFZ (Potsdam, Germany), GGM02S \citep{ggm02} and GGM03S \citep{ggm03} from CSR (Austin, Texas), ITG-Grace02s \citep{ITG} and ITG-Grace03s
\citep{itggrace03s} from IGG (Bonn, Germany), JEM01-RL03B from JPL (NASA, USA) and AIUB-GRACE01S \citep{aiub} from AIUB (Switzerland).
Note that this approach was explicitly followed also by \citet{Ciu96} with the JGM3 and GEMT-2 models.
In Table \ref{tavola1}--Table \ref{tavolaAIUB2} we quote both the sum $\sum_{\ell=4}^{20}f_{\ell}$ of the absolute values of the individual mismodelled terms \eqi f_{\ell} =\left|\dot\Omega_{.\ell}^{\rm LAGEOS} + c_1\dot\Omega_{.\ell}^{\rm LAGEOS\ II}\right|\Delta J_{\ell}\eqf (SAV), and the square root of the sum of their squares $\sqrt{\sum_{\ell=4}^{20}f^2_{\ell}}$ (RSS); in both cases we normalized them to the combined Lense-Thirring total precession of 47.8 mas yr$^{-1}$.
\begin{table}[!h]
   \caption{Impact of the mismodelling in the even zonal harmonics on $f_{\ell}=\left|\dot\Omega^{\rm LAGEOS}_{\ell} + c_1\dot\Omega^{\rm LAGEOS\ II}_{.\ell}\right|\Delta J_{\ell},\ \ell=4,\dots,20$, in mas yr$^{-1}$. Recall that $J_{\ell}=-\sqrt{2\ell + 1}\ \overline{C}_{\ell 0}$; for the uncertainty in the even zonals we have taken here the difference $\Delta\overline{C}_{\ell 0}=\left|\overline{C}_{\ell 0}^{\rm (X)}-\overline{C}_{\ell 0}^{\rm (Y)}\right|$ between the model X=GGM02S \protect\citep{ggm02} and the model Y=ITG-Grace02s \protect\citep{ITG}.
   GGM02S is based on 363 days of GRACE-only data   (GPS and intersatellite tracking, neither constraints nor regularization applied)
spread between April 4, 2002 and Dec 31, 2003. The $\sigma$ are formal for both models. $\Delta \overline{C}_{\ell 0}$ are always larger than the linearly added sigmas, apart from   $\ell=12$ and $\ell=18$. Values of $f_{\ell}$ smaller than 0.1 mas yr$^{-1}$ have not been quoted. The Lense-Thirring precession of the combination of \rfr{combi} amounts to 47.8 mas yr$^{-1}$. The percent bias $\delta\mu$ have been computed by normalizing the linear sum of $f_{\ell}, \ell=4,\dots,20$ (SAV) and the square root of the sum of $f_\ell^2, \ell=4,\dots,20$ to the Lense-Thirring combined precessions.
}\label{tavola1}
\begin{tabular}{llll}
\hline\noalign{\smallskip}
$\ell$ & $\Delta\overline{C}_{\ell 0}$ (GGM02S-ITG-Grace02s) & $\sigma_{\rm  X}+\sigma_{\rm Y}$ & $f_{\ell}$  (mas yr$^{-1}$)\\
\noalign{\smallskip}\hline\noalign{\smallskip}
4 & $1.9\times 10^{-11}$ &  $8.7\times 10^{-12}$ & 7.2\\
6 & $2.1\times 10^{-11}$ &  $4.6\times 10^{-12}$ & 4.6\\
8 & $5.7\times 10^{-12}$ &  $2.8\times 10^{-12}$ & 0.2\\
10 & $4.5\times 10^{-12}$ &  $2.0\times 10^{-12}$ & -\\
12 & $1.5\times 10^{-12}$ &  $1.8\times 10^{-12}$ & -\\
14 & $6.6\times 10^{-12}$ &  $1.6\times 10^{-12}$ & -\\
16 & $2.9\times 10^{-12}$ &  $1.6\times 10^{-12}$ & -\\
18 & $1.4\times 10^{-12}$ &  $1.6\times 10^{-12}$ & -\\
20 & $2.0\times 10^{-12}$ &  $1.6\times 10^{-12}$ & -\\

\noalign{\smallskip}\hline\noalign{\smallskip}
 &    $\delta\mu = 25\%$ (SAV) & $\delta\mu = 18\%$ (RSS) &   \\  %
\noalign{\smallskip}\hline\noalign{\smallskip} %
\end{tabular}
\end{table}
%
%
%
%
%
%
%
%
%
%
%
%
%
%
%
%
%
%
%

\begin{table}[!h]
   \caption{Bias due to the mismodelling in the even zonals of the models X=ITG-Grace03s \protect\citep{itggrace03s}, based on GRACE-only accumulated normal equations from data out of September 2002-April 2007 (neither apriori information nor regularization used), and Y=GGM02S \protect\citep{ggm02}.  The $\sigma$ for both models are formal. $\Delta \overline{C}_{\ell 0}$ are always larger than the linearly added sigmas, apart from  $\ell=12$ and $\ell=18$.}\label{tavola11}
\begin{tabular}{llll}
\hline\noalign{\smallskip}
$\ell$ & $\Delta\overline{C}_{\ell 0}$ (ITG-Grace03s-GGM02S) & $\sigma_{\rm  X}+\sigma_{\rm Y}$ & $f_{\ell}$  (mas yr$^{-1}$)\\
\noalign{\smallskip}\hline\noalign{\smallskip}
4 & $2.58\times 10^{-11}$ &  $8.6\times 10^{-12}$ & 9.6\\
6 & $1.39\times 10^{-11}$ &  $4.7\times 10^{-12}$ & 3.1\\
8 & $5.6\times 10^{-12}$ &  $2.9\times 10^{-12}$ & 0.2\\
10 & $1.03\times 10^{-11}$ &  $2\times 10^{-12}$ & -\\
12 & $7\times 10^{-13}$ &  $1.8\times 10^{-12}$ & -\\
14 & $7.3\times 10^{-12}$ &  $1.6\times 10^{-12}$ & -\\
16 & $2.6\times 10^{-12}$ &  $1.6\times 10^{-12}$ & -\\
18 & $8\times 10^{-13}$ &  $1.6\times 10^{-12}$ & -\\
20 & $2.4\times 10^{-12}$ &  $1.6\times 10^{-12}$ & -\\

\noalign{\smallskip}\hline\noalign{\smallskip}
&    $\delta\mu = 27\%$ (SAV) & $\delta\mu = 21\%$ (RSS) &   \\  %
\noalign{\smallskip}\hline\noalign{\smallskip} %
\end{tabular}
\end{table}
\begin{table}[!h]
   \caption{Bias due to the mismodelling in the even zonals of the models  X = GGM02S \protect\citep{ggm02} and Y = GGM03S \protect\citep{ggm03} retrieved from data spanning January 2003 to December 2006.
    The $\sigma$ for GGM03S are calibrated. $\Delta \overline{C}_{\ell 0}$ are larger than the linearly added sigmas for $\ell = 4,6$. (The other zonals are of no concern)}\label{tavola03S}
\begin{tabular}{llll}
\hline\noalign{\smallskip}
$\ell$ & $\Delta\overline{C}_{\ell 0}$ (GGM02S-GGM03S) & $\sigma_{\rm  X}+\sigma_{\rm Y}$ & $f_{\ell}$  (mas yr$^{-1}$)\\
\noalign{\smallskip}\hline\noalign{\smallskip}
4 & $1.87\times 10^{-11}$ &  $1.25\times 10^{-11}$ & 6.9\\
6 & $1.96\times 10^{-11}$ &  $6.7\times 10^{-12}$ & 4.2\\
8 & $3.8\times 10^{-12}$ &  $4.3\times 10^{-12}$ & 0.1\\
10 & $8.9\times 10^{-12}$ &  $2.8\times 10^{-12}$ & 0.1\\
12 & $6\times 10^{-13}$ &  $2.4\times 10^{-12}$ & -\\
14 & $6.6\times 10^{-12}$ &  $2.1\times 10^{-12}$ & -\\
16 & $2.1\times 10^{-12}$ &  $2.0\times 10^{-12}$ & -\\
18 & $1.8\times 10^{-12}$ &  $2.0\times 10^{-12}$ & -\\
20 & $2.2\times 10^{-12}$ &  $1.9\times 10^{-12}$ & -\\

\noalign{\smallskip}\hline\noalign{\smallskip}
&    $\delta\mu = 24\%$ (SAV) & $\delta\mu = 17\%$ (RSS) &   \\  %
\noalign{\smallskip}\hline\noalign{\smallskip} %
\end{tabular}
\end{table}
\begin{table}[!h]
   \caption{Bias due to the mismodelling in the even zonals of the models  X = EIGEN-GRACE02S \protect\citep{eigengrace02s} and Y = GGM03S \protect\citep{ggm03}.
    The $\sigma$ for both models are calibrated. $\Delta \overline{C}_{\ell 0}$ are always larger than the linearly added sigmas apart from $\ell = 14,18$.}\label{tavola033S}
 \begin{tabular}{llll}
\hline\noalign{\smallskip}
$\ell$ & $\Delta\overline{C}_{\ell 0}$ (EIGEN-GRACE02S-GGM03S) & $\sigma_{\rm  X}+\sigma_{\rm Y}$ & $f_{\ell}$  (mas yr$^{-1}$)\\
\noalign{\smallskip}\hline\noalign{\smallskip}
4 & $2.00\times 10^{-11}$ &  $8.1\times 10^{-12}$ & 7.4\\
6 & $2.92\times 10^{-11}$ &  $4.3\times 10^{-12}$ & 6.3\\
8 & $1.05\times 10^{-11}$ &  $3.0\times 10^{-12}$ & 0.4\\
10 & $7.8\times 10^{-12}$ &  $2.9\times 10^{-12}$ & 0.1\\
12 & $3.9\times 10^{-12}$ &  $1.8\times 10^{-12}$ & -\\
14 & $5\times 10^{-13}$ &  $1.7\times 10^{-12}$ & -\\
16 & $1.7\times 10^{-12}$ &  $1.4\times 10^{-12}$ & -\\
18 & $2\times 10^{-13}$ &  $1.4\times 10^{-12}$ & -\\
20 & $2.5\times 10^{-12}$ &  $1.4\times 10^{-12}$ & -\\

\noalign{\smallskip}\hline\noalign{\smallskip}
&    $\delta\mu = 30\%$ (SAV) & $\delta\mu = 20\%$ (RSS) &   \\  %
\noalign{\smallskip}\hline\noalign{\smallskip}
\end{tabular}
\end{table}
\begin{table}[!h]
   \caption{Bias due to the mismodelling in the even zonals of the models  X = JEM01-RL03B, based on 49 months of GRACE-only data, and Y = GGM03S \protect\citep{ggm03}.
    The $\sigma$ for GGM03S are calibrated. $\Delta \overline{C}_{\ell 0}$ are always larger than the linearly added sigmas apart from $\ell = 16$.}\label{tavolaJEM1}
\begin{tabular}{llll}
\hline\noalign{\smallskip}
$\ell$ & $\Delta\overline{C}_{\ell 0}$ (JEM01-RL03B-GGM03S) & $\sigma_{\rm  X}+\sigma_{\rm Y}$ & $f_{\ell}$  (mas yr$^{-1}$)\\
\noalign{\smallskip}\hline\noalign{\smallskip}
4 & $1.97\times 10^{-11}$ &  $4.3\times 10^{-12}$ & 7.3\\
6 & $2.7\times 10^{-12}$ &  $2.3\times 10^{-12}$ & 0.6\\
8 & $1.7\times 10^{-12}$ &  $1.6\times 10^{-12}$ & -\\
10 & $2.3\times 10^{-12}$ &  $8\times 10^{-13}$ & -\\
12 & $7\times 10^{-13}$ &  $7\times 10^{-13}$ & -\\
14 & $1.0\times 10^{-12}$ &  $6\times 10^{-13}$ & -\\
16 & $2\times 10^{-13}$ &  $5\times 10^{-13}$ & -\\
18 & $7\times 10^{-13}$ &  $5\times 10^{-13}$ & -\\
20 & $5\times 10^{-13}$ &  $4\times 10^{-13}$ & -\\

\noalign{\smallskip}\hline\noalign{\smallskip}
&    $\delta\mu = 17\%$ (SAV) & $\delta\mu = 15\%$ (RSS) &   \\  %
\noalign{\smallskip}\hline\noalign{\smallskip} %
\end{tabular}
\end{table}
\begin{table}[!h]
   \caption{Bias due to the mismodelling in the even zonals of the models  X = JEM01-RL03B and Y = ITG-Grace03s \protect\citep{itggrace03s}.
    The $\sigma$ for ITG-Grace03s are formal. $\Delta \overline{C}_{\ell 0}$ are always larger than the linearly added sigmas.}\label{tavolaJEM2}
\begin{tabular}{llll}
\hline\noalign{\smallskip}
$\ell$ & $\Delta\overline{C}_{\ell 0}$ (JEM01-RL03B-ITG-Grace03s) & $\sigma_{\rm  X}+\sigma_{\rm Y}$ & $f_{\ell}$  (mas yr$^{-1}$)\\
\noalign{\smallskip}\hline\noalign{\smallskip}
4 & $2.68\times 10^{-11}$ &  $4\times 10^{-13}$ & 9.9\\
6 & $3.0\times 10^{-12}$ &  $2\times 10^{-13}$ & 0.6\\
8 & $3.4\times 10^{-12}$ &  $1\times 10^{-13}$ & 0.1\\
10 & $3.6\times 10^{-12}$ &  $1\times 10^{-13}$ & -\\
12 & $6\times 10^{-13}$ &  $9\times 10^{-14}$ & -\\
14 & $1.7\times 10^{-12}$ &  $9\times 10^{-14}$ & -\\
16 & $4\times 10^{-13}$ &  $8\times 10^{-14}$ & -\\
18 & $4\times 10^{-13}$ &  $8\times 10^{-14}$ & -\\
20 & $7\times 10^{-13}$ &  $8\times 10^{-14}$ & -\\

\noalign{\smallskip}\hline\noalign{\smallskip}
&    $\delta\mu = 22\%$ (SAV) & $\delta\mu = 10\%$ (RSS) &   \\  %
\noalign{\smallskip}\hline\noalign{\smallskip}
\end{tabular}
\end{table}
\begin{table}[!h]
   \caption{Aliasing effect of the mismodelling in the even zonal harmonics estimated in the X=ITG-Grace03s \protect\citep{itggrace03s} and the Y=EIGEN-GRACE02S \protect\citep{eigengrace02s} models.  The covariance matrix $\sigma$ for ITG-Grace03s are formal, while the ones of EIGEN-GRACE02S are calibrated. $\Delta \overline{C}_{\ell 0}$ are larger than the linearly added sigmas for $\ell =4,...,20$, apart from $\ell=18$. }\label{tavola7}
\begin{tabular}{llll}
\hline\noalign{\smallskip}
$\ell$ & $\Delta\overline{C}_{\ell 0}$ (ITG-Grace03s-EIGEN-GRACE02S) & $\sigma_{\rm  X}+\sigma_{\rm Y}$ & $f_{\ell}$  (mas yr$^{-1}$)\\
\noalign{\smallskip}\hline\noalign{\smallskip}
4 & $2.72\times 10^{-11}$ &  $3.9\times 10^{-12}$ & 10.1\\
6 & $2.35\times 10^{-11}$ &  $2.0\times 10^{-12}$ & 5.1\\
8 & $1.23\times 10^{-11}$ &  $1.5\times 10^{-12}$ & 0.4\\
10 & $9.2\times 10^{-12}$ &  $2.1\times 10^{-12}$ & 0.1\\
12 & $4.1\times 10^{-12}$ &  $1.2\times 10^{-12}$ & -\\
14 & $5.8\times 10^{-12}$ &  $1.2\times 10^{-12}$ & -\\
16 & $3.4\times 10^{-12}$ &  $9\times 10^{-13}$ & -\\
18 & $5\times 10^{-13}$ &  $1.0\times 10^{-12}$ & -\\
20 & $1.8\times 10^{-12}$ &  $1.1\times 10^{-12}$ & -\\

\noalign{\smallskip}\hline\noalign{\smallskip}
&    $\delta\mu = 37\%$ (SAV) & $\delta\mu = 24\%$ (RSS) &   \\  %
\noalign{\smallskip}\hline\noalign{\smallskip} %
\end{tabular}

\end{table}
%
%
%
%
           %
 %
%
%
%
%
%
%
%
\begin{table}[!h]
   \caption{Bias due to the mismodelling in the even zonals of the models  X = JEM01-RL03B, based on 49 months of GRACE-only data, and Y = AIUB-GRACE01S \protect\citep{aiub}. The latter one was obtained from GPS satellite-to-satellite tracking data and K-band range-rate data out of the
period January 2003 to December 2003 using the Celestial Mechanics Approach.
No accelerometer data, no de-aliasing products, and no regularisation was
applied.
    The $\sigma$ for AIUB-GRACE01S are formal.
    $\Delta \overline{C}_{\ell 0}$ are always larger than the linearly added sigmas.
    }\label{tavolaAIUB1}
\begin{tabular}{llll}
\hline\noalign{\smallskip}
$\ell$ & $\Delta\overline{C}_{\ell 0}$ (JEM01-RL03B$-$AIUB-GRACE01S) & $\sigma_{\rm  X}+\sigma_{\rm Y}$ & $f_{\ell}$  (mas yr$^{-1}$)\\
\noalign{\smallskip}\hline\noalign{\smallskip}
4 & $2.95\times 10^{-11}$ &  $2.1\times 10^{-12}$ & 11\\
6 & $3.5\times 10^{-12}$  &  $1.3\times 10^{-12}$ & 0.8\\
8 & $2.14\times 10^{-11}$ &  $5\times 10^{-13}$ & 0.7\\
10 & $4.8\times 10^{-12}$ &  $5\times 10^{-13}$ & -\\
12 & $4.2\times 10^{-12}$ &  $5\times 10^{-13}$ & -\\
14 & $3.6\times 10^{-12}$ &  $5\times 10^{-13}$ & -\\
16 & $8\times 10^{-13}$ &    $5\times 10^{-13}$ & -\\
18 & $7\times 10^{-13}$  &    $5\times 10^{-13}$ & -\\
20 & $1.0\times 10^{-12}$ &    $5\times 10^{-13}$ & -\\

\noalign{\smallskip}\hline\noalign{\smallskip}
&   $\delta\mu = 26\%$ (SAV)&    $\delta\mu = 23\%$ (RSS)& \\  %
\noalign{\smallskip}\hline\noalign{\smallskip} %
\end{tabular}
\end{table}
\begin{table}[!h]
   \caption{Bias due to the mismodelling in the even zonals of the models  X = EIGEN-GRACE02S \protect\citep{eigengrace02s} and Y = AIUB-GRACE01S \protect\citep{aiub}. The $\sigma$ for AIUB-GRACE01S are formal, while those of EIGEN-GRACE02S are calibrated.
    $\Delta \overline{C}_{\ell 0}$ are  larger than the linearly added sigmas for $\ell=4,6,8,16$.
    }\label{tavolaAIUB2}
\begin{tabular}{llll}
\hline\noalign{\smallskip}
$\ell$ & $\Delta\overline{C}_{\ell 0}$ (EIGEN-GRACE02S$-$AIUB-GRACE01S) & $\sigma_{\rm  X}+\sigma_{\rm Y}$ & $f_{\ell}$  (mas yr$^{-1}$)\\
\noalign{\smallskip}\hline\noalign{\smallskip}
4 & $2.98\times 10^{-11}$ &  $6.0\times 10^{-12}$ & 11.1\\
6 & $2.29\times 10^{-11}$  &  $3.3\times 10^{-12}$ & 5.0\\
8 & $1.26\times 10^{-11}$ &  $1.9\times 10^{-12}$ & 0.4\\
10 & $6\times 10^{-13}$ &  $2.5\times 10^{-12}$ & -\\
12 & $5\times 10^{-13}$ &  $1.6\times 10^{-12}$ & -\\
14 & $5\times 10^{-13}$ &  $1.6\times 10^{-12}$ & -\\
16 & $2.9\times 10^{-12}$ &    $1.4\times 10^{-12}$ & -\\
18 & $6\times 10^{-13}$  &    $1.4\times 10^{-12}$ & -\\
20 & $2\times 10^{-13}$ &    $1.5\times 10^{-12}$ & -\\

\noalign{\smallskip}\hline\noalign{\smallskip}
&   $\delta\mu = 34\%$ (SAV)&    $\delta\mu = 25\%$ (RSS)& \\  %
\noalign{\smallskip}\hline\noalign{\smallskip} %
\end{tabular}
\end{table}

The systematic bias evaluated with a more realistic approach is about 3 to 4 times larger than one can obtain by only using this or that particular model. The scatter is still quite large and far from the $5-10\%$ claimed in \citep{Ciu04}. In particular, it appears that $J_4$, $J_6$, and to a lesser extent $J_8$, which are just the most relevant zonals for us because of their impact on the combination of \rfr{combi}, are the most uncertain ones, with discrepancies $\Delta J_{\ell}$ between different models, in general, larger than the sum of their sigmas $\sigma_{J_{\ell}}$, calibrated or not.  

Another way to evaluate the uncertainty in the LAGEOS-LAGEOS II node test may consist of computing the nominal values of the total combined precessions for different models  and comparing them, i.e. by taking
\eqi \left|\sum_{\ell = 4}\left(\dot\Omega_{.\ell}^{\rm LAGEOS} + c_1\dot\Omega_{.\ell}^{\rm LAGEOS\ II}\right) [J_{\ell}(\rm  X)-J_{\ell}(\rm Y)]\right|.\eqf The results are shown in
Table \ref{tavolaa}.
 \begin{table}[!h]\caption{ Systematic uncertainty $\delta\mu$ in the LAGEOS-LAGEOS II test evaluated by taking the absolute value of the difference between the nominal values of the total combined node precessions due to the even zonals for different models X and Y, i.e. $\left|
 \dot\Omega^{\rm geopot}({\rm X})-\dot\Omega^{\rm geopot}({\rm Y})\right|$.}\label{tavolaa}

\begin{tabular}{ll}
\hline\noalign{\smallskip}
Models compared & $\delta\mu$  \\
\noalign{\smallskip}\hline\noalign{\smallskip}
AIUB-GRACE01S$-$JEM01-RL03B & $20\%$\\
%
%
AIUB-GRACE01S$-$GGM02S & $27\%$\\
AIUB-GRACE01S$-$GGM03S & $3\%$\\
AIUB-GRACE01S$-$ITG-Grace02 & $2\%$\\
AIUB-GRACE01S$-$ITG-Grace03 & $0.1\%$\\
%
%
AIUB-GRACE01S$-$EIGEN-GRACE02S & $33\%$\\
%
%
%
%
%
%
%
%
%
JEM01-RL03B$-$GGM02S & $7\%$\\
JEM01-RL03B$-$GGM03S & $17\%$\\
JEM01-RL03B$-$ITG-Grace02 & $18\%$\\
JEM01-RL03B$-$ITG-Grace03s & $20\%$\\
%
%
JEM01-RL03B$-$EIGEN-GRACE02S & $13\%$\\
%
%
GGM02S$-$GGM03S & $24\%$\\
GGM02S$-$ITG-Grace02& $25\%$\\
GGM02S$-$ITG-Grace03s& $27\%$\\
%
%
GGM02S$-$EIGEN-GRACE02S & $6\%$\\
%
%
GGM03S$-$ITG-Grace02 & $1\%$\\
GGM03S$-$ITG-Grace03s & $3\%$\\
%
%
GGM03S$-$EIGEN-GRACE02S & $30\%$\\
%
%
ITG-Grace02$-$ITG-Grace03s & $2\%$\\
%
%
ITG-Grace02$-$EIGEN-GRACE02S & $31\%$\\
%
%
%
ITG-Grace03s$-$EIGEN-GRACE02S & $33\%$\\
%
%
%
%
%
\noalign{\smallskip}\hline

\end{tabular}

\end{table}

A different approach that could be followed to take into account the scatter among the various solutions consists in computing mean and standard deviation of the entire set of values of the even zonals for the models considered so far, degree by degree, and    taking the standard deviations as representative of the uncertainties $\delta J_{\ell}, \ell = 4,6,8,...$. It yields $\delta\mu = 15\%$, in agreement with \citet{Ries08}.

It must be recalled that also the further bias due to the cross-coupling between $J_2$ and the orbit inclination, evaluated to be about $9\%$ in \citep{Ior07}, must be added.

\section{A new approach to extract the Lense-Thirring signature from the data}\lb{approach}
The technique adopted so far in \citep{Ciu04} and \citep{Ries08} to extract the gravitomagentic signal from the LAGEOS and LAGEOS II data  is described in detail in
\citep{LucBal06,Luc07}. In both the approaches the Lense-Thirring force is not included in the dynamical force models used to fit the satellites' data. In the data reduction process no dedicated gravitomagnetic parameter is estimated, contrary to, e.g., station coordinates, state vector,
satellites' drag and radiation coefficients $C_D$ and $C_R$, respectively, etc.; its effect is retrieved with a sort of post-post-fit analysis in which the time series of the computed\footnote{The expression ``residuals of the nodes'' is used, strictly speaking, in an improper sense because the Keplerian orbital elements are not directly measured.} ``residuals'' of the nodes with the difference between the orbital elements of consecutive arcs, combined with \rfr{combi}, is fitted with a straight line.

In order to enforce the reliability of the ongoing test it would be desirable to proceed following other approaches as well. For instance, the gravitomagnetic force could be explicitly modelled in terms of a dedicated solve-for parameter (not necessarily the usual PPN $\gamma$ one) to  be estimated in the least-square sense along with all the other parameters usually determined in fitting the LAGEOS-type satellites data, and the resulting correlations among them  could be inspected. Moreover, one could also look at the changes in the values of the complete set of the estimated parameters with and without the Lense-Thirring effect modelled.

A first, tentative step  towards the implementation of a similar strategy with the LAGEOS satellites in terms  of the PPN parameter $\gamma$ has been recently taken by \citet{Poz}.

\section{On the LARES mission}\lb{larez}
The combination that should be used for measuring the Lense-Thirring effect with LAGEOS, LAGEOS II and LARES is  \citep{IorNA}
\eqi \dot\Omega^{\rm LAGEOS}+k_1\dot\Omega^{\rm LAGEOS\ II}+ k_2\dot\Omega^{\rm LARES}.
\lb{combaz}\eqf
The coefficients $k_1$ and $k_2$ entering \rfr{combaz} are defined as
\begin{equation}
\begin{array}{lll}
k_1 = \rp{\cf 2{LARES}\cf4{LAGEOS}-\cf 2{LAGEOS}\cf 4{LARES}}{\cf 2{LAGEOS\ II}\cf 4{LARES}-\cf 2{LARES}\cf 4{LAGEOS\ II}}= 0.3586,\\\\
k_2 =  \rp{\cf 2{LAGEOS}\cf4{LAGEOS\ II}-\cf 2{LAGEOS\ II}\cf 4{LAGEOS}}{\cf 2{LAGEOS\ II}\cf 4{LARES}-\cf 2{LARES}\cf 4{LAGEOS\ II}}= 0.0751.
\end{array}\lb{cofis}
 \end{equation}
The combination \rfr{combaz} cancels out, by construction, the impact of the first two even zonals; we have used $a_{\rm LR}=7828$ km, $i_{\rm LR}=71.5$ deg.
The total Lense-Thirring effect, according to \rfr{combaz} and \rfr{cofis}, amounts to 50.8 mas yr$^{-1}$.
\subsection{A conservative evaluation of the impact of the geopotential on the LARES mission}
The systematic error due to the uncancelled even zonals $J_6, J_8,...$ can be conservatively evaluated as
\eqi\delta\mu\leq \sum_{\ell = 6}\left|\dot\Omega^{\rm LAGEOS}_{.\ell}+k_1\dot\Omega^{\rm LAGEOS\ II}_{.\ell}+ k_2\dot\Omega^{\rm LARES}_{.\ell}\right|\delta J_{\ell}\lb{biass}\eqf

Of crucial importance is how to assess $\delta J_{\ell}$.  By proceeding as in Section \ref{grav} and by using the same models up to degree $\ell = 60$  because of the lower altitude of LARES with respect to LAGEOS and LAGEOS II which brings into play more even zonals, we have the results presented in Table \ref{tavolay}. They have been obtained with the standard and widely used Kaula approach \citep{Kau} in the following way. We, first, calibrated our numerical calculation with the analytical ones performed with the explicit expressions for $\dot\Omega_{.\ell}$ worked out up to $\ell=20$ in \citep{Ior03}; then, after having obtained identical results, we confidently extended our numerical calculation to higher degrees by means of two different softwares.
 \begin{table}[!h]\caption{ Systematic percent uncertainty $\delta\mu$ in the combined Lense-Thirring effect with LAGEOS, LAGEOS II and LARES according to \rfr{biass} and $\delta J_{\ell}= \Delta J_{\ell}$ up to degree $\ell = 60$ for the global Earth's gravity solutions  considered here; the approach by \protect\citep{Kau} has been followed. For LARES we adopted $a_{\rm LR}=7828$ km, $i_{\rm LR}=71.5$ deg, $e_{\rm LR} = 0.0$.}\label{tavolay}

\begin{tabular}{lll}
\hline\noalign{\smallskip}
Models compared ($\delta J_{\ell}=\Delta J_{\ell}$) & $\delta\mu$ (SAV) & $\delta\mu$ (RSS)\\
\noalign{\smallskip}\hline\noalign{\smallskip}
AIUB-GRACE01S$-$JEM01-RL03B & $23\%$ & $16\%$\\
%
%
AIUB-GRACE01S$-$GGM02S & $16\%$ & $8\%$\\
AIUB-GRACE01S$-$GGM03S & $22\%$ & $13\%$\\
AIUB-GRACE01S$-$ITG-Grace02 & $24\%$ & $15\%$\\
AIUB-GRACE01S$-$ITG-Grace03 & $22\%$ & $14\%$\\
%
%
AIUB-GRACE01S$-$EIGEN-GRACE02S & $14\%$ & $7\%$\\
%
%
%
%
%
%
%
%
%
JEM01-RL03B$-$GGM02S & $14\%$ & $9\%$  \\
JEM01-RL03B$-$GGM03S & $5\%$ & $3\%$  \\
JEM01-RL03B$-$ITG-Grace02 & $4\%$ & $2\%$  \\
JEM01-RL03B$-$ITG-Grace03s & $5\%$ & $2\%$  \\
%
%
JEM01-RL03B$-$EIGEN-GRACE02S & $26\%$ & $15\%$  \\
%
%
GGM02S$-$GGM03S & $13\%$ & $7\%$  \\
GGM02S$-$ITG-Grace02& $16\%$ & $8\%$  \\
GGM02S$-$ITG-Grace03s& $14\%$ & $7\%$  \\
%
%
GGM02S$-$EIGEN-GRACE02S & $14\%$ & $7\%$  \\
%
%
GGM03S$-$ITG-Grace02 & $3\%$ & $2\%$  \\
GGM03S$-$ITG-Grace03s & $2\%$ & $0.5\%$  \\
%
%
GGM03S$-$EIGEN-GRACE02S & $24\%$ & $13\%$  \\
%
%
ITG-Grace02$-$ITG-Grace03s & $3\%$ & $2\%$  \\
%
%
ITG-Grace02$-$EIGEN-GRACE02S & $25\%$ & $14\%$  \\
%
%
%
ITG-Grace03s$-$EIGEN-GRACE02S & $24\%$ & $13\%$  \\
%
%
%
%
%
\noalign{\smallskip}\hline

\end{tabular}

\end{table}

It must be stressed that they may be still optimistic: indeed, computations for $\ell > 60$ become unreliable because of numerical instability of the results.

In Table \ref{tavolax} we repeat the calculation by using for $\delta J_{\ell}$ the covariance matrix sigmas $\sigma_{J_{\ell}}$; also in this case we use the approach by \citet{Kau} up to degree $\ell = 60$.
 \begin{table}[!h]\caption{ Systematic percent uncertainty $\delta\mu$ in the combined Lense-Thirring effect with LAGEOS, LAGEOS II and LARES according to \rfr{biass} and $\delta J_{\ell}= \sigma_{J_{\ell}}$ up to degree $\ell = 60$ for the global Earth's gravity solutions  considered here; the approach by \protect\citep{Kau} has been followed. For LARES we adopted $a_{\rm LR}=7828$ km, $i_{\rm LR}=71.5$ deg, $e_{\rm LR} = 0.0$.}\label{tavolax}

\begin{tabular}{lll}
\hline\noalign{\smallskip}
Model ($\delta J_{\ell}=\sigma_{\ell}$) & $\delta\mu$ (SAV) & $\delta\mu$ (RSS)\\
\noalign{\smallskip}\hline\noalign{\smallskip}%
AIUB-GRACE01S (formal) & $11\%$ & $9\%$\\
%
 %
JEM01-RL03B & $1\%$ & $0.9\%$\\
GGM03S (calibrated) & $5\%$ & $4\%$\\
GGM02S (formal) & $20\%$ & $15\%$\\
ITG-Grace03s (formal) & $0.3\%$ & $0.2\%$\\
ITG-Grace02s (formal) & $0.4\%$ & $0.2\%$\\
%
%
EIGEN-GRACE02S (calibrated) & $21\%$ & $17\%$ \\
 \noalign{\smallskip}\hline

\end{tabular}

\end{table}

If, instead, one assumes $\delta J_{\ell}=s_{\ell},\ \ell=2,4,6,...$ i.e., the standard deviations of the sets of all the best estimates of $J_{\ell}$ for the models considered here the systematic bias, up to $\ell=60$, amounts to $12\%$ (SAV) and $6\%$ (RSS). Again, also this result may turn out to be optimistic for the same reasons as before.


It must be pointed out that the evaluations presented here rely upon calculations of the coefficients $\dot\Omega_{.\ell}$ performed with the well known standard approach by Kaula \citep{Kau}; it would be important to try to follow also different computational strategies in order to test them.
\subsection{The impact of some non-gravitational perturbations}
It is worthwhile noting that also the impact of the subtle non-gravitational perturbations will be different with respect to the original proposal because LARES will fly in a  lower orbit and its thermal behavior will  probably be different with respect to  LAGEOS and LAGEOS II.  The reduction of the impact of the thermal accelerations, like the Yarkovsky-Schach effects, should have  been reached with two concentric spheres. However, as explained by \citet{Andres}, this solution will increase the floating potential of LARES because of the much higher electrical resistivity and, thus, the perturbative effects produced by the charged particle drag. Moreover, the atmospheric drag will increase also because of the lower orbit of the satellite, both in its neutral and charged components.  Indeed, although it does not affect directly the node $\Omega$, it induces a secular decrease of the inclination $i$ of a LAGEOS-like satellite \citep{Mil87} which translates into a further bias for the node itself according to
\eqi\delta\dot\Omega_{\rm drag}=\rp{3}{2}n\left(\rp{R}{a}\right)^2 \rp{\sin i\ J_2}{(1-e^2)^2}\delta i,\eqf in which $\delta i$ accounts not only for the measurement errors in the inclination, but also for any unmodelled/mismodelled dynamical effect on it. According to \citep{Iordrag}, the secular decrease for LARES would amount to \eqi \left\langle\dert i t\right\rangle_{\rm LR}\approx -0.6\ {\rm mas}\ {\rm yr}^{-1}\eqf yielding a systematic uncertainty in the Lense-Thirring signal of \rfr{combaz} of about $3-9\%$ yr$^{-1}$. An analogous indirect node effect via the inclination could be induced by the thermal Yarkovski-Rubincam force as well \citep{Iordrag}.
Also the Earth's albedo, with its anisotropic components, may have a non-negligible  effect.

Let us point out the following issue as well. At present, it is not yet clear how the data of LAGEOS, LAGEOS II and LARES will be finally used by the proponent team in order to try to detect the Lense-Thirring effect. This could turn out to be a non-trivial matter because of the non-gravitational perturbations. Indeed, if, for instance, a combination\footnote{The impact of the geopotential is, by construction, unaffected with respect to the combination of \rfr{combaz}.}
\eqi\dot\Omega^{\rm LARES}+h_1\dot\Omega^{\rm LAGEOS}+ h_2\dot\Omega^{\rm LAGEOS\ II}\lb{altr}\eqf was adopted instead of that of \rfr{combaz}, the coefficients of the nodes of LAGEOS and LAGEOS II, in view of the lower altitude of LARES, would be
 \begin{equation}
\begin{array}{lll}
h_1 = \rp{\cf 2{LAGEOS\ II}\cf4{LARES}-\cf 2{LARES}\cf 4{LAGEOS\ II}}{\cf 2{LARES}\cf 4{LAGEOS\ II}-\cf 2{LAGEOS\ II}\cf 4{LAGEOS}}= 13.3215,\\\\
h_2 =  \rp{\cf 2{LARES}\cf4{LAGEOS}-\cf 2{LAGEOS}\cf 4{LARES}}{\cf 2{LAGEOS}\cf 4{LAGEOS\ II}-\cf 2{LAGEOS\ II}\cf 4{LAGEOS}}= 4.7744.
\end{array}\lb{cofcaz}
 \end{equation}
and the combined Lense-Thirring signal would amount to 676.8 mas yr$^{-1}$.
As a consequence, the direct and indirect effects of the non-gravitational\footnote{The same may hold also for time-dependent gravitational perturbations affecting the nodes of LAGEOS and LAGEOS II, like the tides.} perturbations on the nodes of LAGEOS and LAGEOS II would be enhanced by such larger coefficients and this may yield a degradation of the total obtainable accuracy.

\section{Conclusions}
The so far published evaluations of the total systematic uncertainty induced by the even zonal harmonics of the geopotential in the Lense-Thirring test with the combined nodes of the SLR LAGEOS and LAGEOS II satellites  are  likely optimistic. Indeed, they are all based on the use of the covariance sigmas, more or less reliably calibrated, of the covariance matrices of various Earth gravity model solutions used one at a time separately in such a way that the model X yields an error of $x\%$, the model Y yields an error $y\%$, etc. Instead,  comparing the estimated values of the even zonals for different pairs of models   allows for a more conservative evaluation of the real uncertainties in our knowledge of the static part of the geopotential. As a consequence, the uncertainty in the Lense-Thirring signal is about $3-4$ times larger than the figures so far claimed ($5-10\%$), amounting to various tens percent ($37\%$ for the pair EIGEN-GRACE02S and ITG-GRACE03s, about $25-30\%$ for the other most recent GRACE-based solutions).

Concerning the extraction of the Lense-Thirring signal from the data of the LAGEOS-type satellites, different approaches with respect to the one followed so far should be implemented in order to do something really new. For instance, the gravitomagnetic force should be explicitly included in the dynamical force models of the LAGEOS satellites and an ad-hoc parameter should be estimated in the least-square sense in addition to those determined so far without modelling the Lense-Thirring effect. Moreover, also the variation of the values of all the other estimated parameters with and without the gravitomagnetic force modelled should be inspected along with their mutual correlations.

  Applying the strategy of the difference of the estimated even zonals to the ongoing LARES mission shows that reaching a $1\%$ measurement of the Lense-Thirring effect with LAGEOS, LAGEOS II and LARES maybe difficult. Indeed, since LARES will orbit at a lower altitude with respect to the LAGEOS satellites, more even zonal harmonics are to be taken into account. Assessing realistically their impact is neither easy nor unambiguous. Straightforward calculations up to degree $\ell = 60$ with the standard Kaula's approach yield errors as large as some tens percent; the same holds if the sigmas of the covariance matrices of several global Earth's gravity models are used. Such an important point certainly deserves great attention. Another issue which may potentially degrade the expected accuracy is the impact of some non-gravitational perturbations which would have a non-negligible effect on LARES  because of its lower orbit. In particular, the secular decrease of the inclination of the new spacecraft due to the neutral and charged atmospheric drag induces an indirect bias in the node precessions by the even zonals which, in the case of LARES, should be of the order of $\approx 3-9\%$ yr$^{-1}$.

\begin{acknowledgements}
I  thank  M.C.E. Huber, R.A. Treumann and the entire staff of ISSI for the organization of the exquisite workshop which I had the pleasure and the honor to attend. I am grateful to D. Barbagallo (ESA-ESRIN) for the information on the LARES orbital configuration. I acknowledge the financial support of INFN-Sezione di Pisa and ISSI.
\end{acknowledgements}

\end{document}